\documentclass[12pt]{article}

\usepackage{amsmath,amssymb}

\makeatletter

\newcommand{\re}{\mathbb{R}}

\@addtoreset{equation}{section}
\makeatother

\begin{document}

\thispagestyle{empty}

\setcounter{page}{0}

\mbox{}

\begin{center} {\bf \Large  The Action for Twisted Self-Duality}

\vspace{1.6cm}

Claudio Bunster$^{1}$ and Marc Henneaux$^{1,2}$

\footnotesize
\vspace{.6 cm}

${}^1${\em Centro de Estudios Cient\'{\i}ficos (CECs), Casilla 1469, Valdivia, Chile}

\vspace{.1cm}

${}^2${\em Universit\'e Libre de Bruxelles and International Solvay Institutes, ULB-Campus Plaine CP231, B-1050 Brussels, Belgium} \\



\vspace {15mm}

\end{center}
\centerline{\bf Abstract}
\vspace{.6cm}
One may write the Maxwell equations in terms of  two gauge potentials, one electric and one magnetic, by demanding that their field strengths should be dual to each other. This requirement is the condition of twisted self-duality. It can be extended to  p-forms in spacetime of D dimensions, and it survives the introduction of a variety of couplings among  forms of different rank, and also to spinor and scalar fields, which emerge naturally from supergravity. In this paper we provide a systematic derivation of the action principle, whose equations of motion are the condition of twisted self-duality. The derivation starts from the standard Maxwell action,  extended to include  the aforementioned couplings,  and proceeds via the Hamiltonian formalism through the resolution of Gauss' law. In the pure Maxwell case we recover in this way an action that had been  postulated by other authors,  through an ansatz based on an action given earlier  by us for untwisted self-duality. When Chern-Simons couplings are included, our action is, however, new. The derivation from the standard extended Maxwell action implies of course that the theory is Lorentz-invariant and can be locally coupled to gravity.   Nevertherless we include a direct compact Hamiltonian proof of these properties, which is based on the surface-deformation algebra. The symmetry in the dependence of the action on the electric and magnetic variables is manifest, since they appear as canonical conjugates.  Spacetime covariance, although present, is not manifest. 
\vspace{.8cm}
\noindent

\newpage

\section{Introduction}
\setcounter{equation}{0}

The symmetry between electricity and magnetism is a fascinating subject. It  originated in the  Maxwell equations, but it has shown a remarkable resilience in front of further developments.  It  survived when spacetime was liberated from the requirement of being four dimensional and also when the door was opened for $p$-forms of an arbitrary rank to come in, as generalizations of the $1$-form of  the Maxwell theory. Today this electric-magnetic ``duality principle" permeates our thinking in supergravity and string theory. 

The duality principle leads naturally to a reformulation of the Maxwell equations, and also of its generalizations mentioned above.  One regards the Maxwell equations as  the conditions for the existence of the usual ``electric potential" $1$-form A and a second  ``magnetic  potential" $1$-form  B. If one demands that the corresponding field strengths (curvature $2$-forms) be the dual of each other one obtains the Maxwell equations. This requirement is called ``twisted self-duality" \cite{Cremmer:1998px}.  The term ``twisted"  is introduced because the  forms are not self-dual, but are rather, as it was just said, dual to each other. If both curvature forms are grouped into a two-component colum, then that colum is related to its dual by an off-diagonal ``twist matrix".
 
When the topology of spacetime is trivial Maxwell equations imply in turn the twisted self Ðduality condition, because every closed two-form is then exact. For non-trivial topologies   additional considerations are needed, which will be addressed in \cite{BHtoappear} (see also \cite{Bekaert:1998yp}).  This paper will be concerned only with the case of trivial topology.

An important motivation for undertaking the present work was the necessity to dispel the widespread misconception that twisted (and untwisted) self-duality can only be discussed at the level of the equations of motion.  

This misconception, which would impede the quantum implementation of duality, has been quite resilient in spite of the fact that the duality invariance of the Maxwell action in four dimensions was already proven in \cite{Deser:1976iy} and that the action for untwisted self-duality was given in reference \cite{Henneaux:1988gg} for chiral $p$-forms in $2p+2$ dimensions.  The action of \cite{Henneaux:1988gg} was then used in \cite{ScSe} as the starting point to arrive at an action for twisted self-duality.

The theme of this paper is a systematic derivation of the action for twisted self-duality from the Maxwell action with Chern-Simons and other $p$-form couplings.  For the pure Maxwell case, the action that we find coincides with that of \cite{ScSe}. When Chern-Simons couplings are included, our action is, however, new.

The action that we deal with is local in space and time and it is quadratic in the fields for the free theories.  It is Lorentz invariant and it can also be locally coupled to the gravitational field. The symmetry in the dependence of the action on the electric and magnetic  variables is manifest, since they appear as canonical conjugates.  Spacetime covariance, although present, is not manifest.

The non-manifest character of spacetime covariance is in sharp constrast with the manifest validity of the duality principle. It  would appear  therefore that in order to spell out the consequences of the duality principle one has to necessarily relegate spacetime covariance to a lesser role. This feature was already encountered in the past in the demonstration of  off-shell duality invariance in a variety of contexts \cite{Deser:1976iy,Deser:1997mz,Bunster:2011aw}, including linearized gravity \cite{Henneaux:2004jw,Julia:2005ze}, in spite of the intimate connection of the latter with spacetime covariance.  It is also present in the action for  chiral bosons \cite{Floreanini:1987as} and self-dual $p$-forms \cite{Henneaux:1988gg}  and was particularly emphasized in \cite{ScSe}. One cannot help but feel that this is an important lesson for the investigations of more general ``hidden symmetries" \cite{Hidden}. Although non-manifest, the spacetime covariance of the action may be proven directly in the present formulation, and in a compact manner, by verifying that the energy and momentum densities satisfy the algebra of surface deformations \cite{Dirac:1962aa,Schwinger:1963xx,Teitelboim:1972vw}.

Other actions have been proposed \cite{Tonin,Bandos,Dall} that are manifestly duality and Lorentz invariant. These actions contain additional fields and additional gauge symmetries. They are non-polynomial even when the interactions are switched off. To get a tractable action, one must fix the new gauge symmetry in a way that breaks Lorentz invariance.  In particular, for the case of a $3$-form with Chern-Simons couplings in eleven dimensions, our action coincides with the one given earlier in \cite{Bandos} when the additional gauge freedom is fixed in a very simple form. 

The situation here is strongly reminiscent of that encountered by ourselves quite a way back, when we developped a Lorentz-invariant formulation of the Hamiltonian dynamics of the superparticle \cite{Brink}. We introduced then extra gauge variables, and concluded that the result was ``rather involved".  Again in that case, the non-manifest Lorentz invariant formulation remained by far the simplest one.

The paper is organized  so as to go go through a number of cases of increasing complexity, treating in detail the simplest of them and then just indicating the results for the more complicated ones. This we do for the sake of focusing on the central point without being distracted by unessential technical burdens.
Thus, Section 2 is devoted to the implementation of twisted self-duality for a  single Maxwell p-form, in  D spacetime dimensions.   We focus on the case of a $1$-form and then indicate the results for a general  $p$. It is explained how the marginal cases $p=0$  and $p =D-2$ fit into the scheme.

The starting point is the Hamiltonian formulation of the standard Maxwell action in tems of a $p$-form potential A, which we call the ``purely electric formulation".  The key step, first devised in \cite{Deser:1976iy}, is solving its Gauss law without going to the reduced phase space, i.e., without fixing the gauge. Since the Gauss law is the vanishing of the divergence of a local vector density , its solution automatically brings in  a $(D-p-2)$-form $B$, which is the magnetic potential. The desired ``electric-magnetic action" is then obtained by introducing the solution of the Gauss constraint of the the original  purely electric action back into it.  Thus the fact that the Gauss constraint is a local divergence is far from being a technicality. It is, rather  a profound manifestation of the duality principle.

Section 3 is devoted to the inclusion of a Chern-Simons term. There again we analyze in detail the simplest case, that is, $p=1$, $D=3$, and then indicate explicitly the results for the generalization to  $p=3$, $D=11$, which is of special interest because it arises in supergravity. The procedure applies however quite generally, since the  Gauss constraint is a divergence for all cases when a Chern-Simons form can be written.

The next step in increasing complexity is taken in section 4, where we show that our procedure can be applied to the coupling among a $1$-form and a $2$-form  that arises in Einstein-Maxwell supergravity in ten dimensions, and indicate its generalization to couplings between several $p$-forms of different degrees. We also remark that the procedure can be applied straightforwardly to Pauli couplings to spinors and to couplings to uncharged scalars.

{} Finally section 5 is devoted to concluding remarks.

\section{Twisted Self-Duality for a  Maxwell $p$-Form in $D$ Spacetime Dimensions}
\label{SinglepForm}
\setcounter{equation}{0}

\subsection{Twisted Self-Duality}

{}For a $p$-form in $D$ spacetime dimensions, there exists a straightforward generalization of the Maxwell action,
\begin{equation}
S[A_{\lambda_1\cdots \lambda_p}] = \int d^{D}x\left( - \frac{1}{2  (p+1)!}  \, F_{\lambda_1\cdots \lambda_{p+1}} F^{\lambda_1\cdots \lambda_{p+1}} \right) \, , \label{ActionpD0}
\end{equation}
with, 
\begin{equation}
F_{\lambda_1\cdots \lambda_{p+1}} = (p+1) \partial_{[\lambda_1} A_{\lambda_2\cdots \lambda_{p+1}]}.
\end{equation}
The square bracket indicates complete antisymmetrization in the enclosed indices,  normalized by dividing by the appropriate factorial so that it is idempotent. 
In terms of forms,
\begin{equation}
F =  dA, \label{Def24ofF}
\end{equation}
with
\begin{equation}
F =  \frac{1}{(p+1)!} F_{\lambda_1\cdots \lambda_{p+1}} dx^{\lambda_1} \wedge \cdots \wedge dx^{\lambda_{p+1}}, \label{Def24ofF'}
\end{equation}
and
\begin{equation}
A = \frac{1}{p!} A_{\lambda_1\cdots \lambda_{p}} dx^{\lambda_1} \wedge \cdots \wedge dx^{\lambda_{p}}.
\end{equation}

The equations of motion obtained by demanding that the action (\ref{ActionpD0}) be stationary with respect to variations of the potential $A$ are,
\begin{equation}
d \;   ^*\hspace{-.05cm} F = 0. \label{EOMstarF}
\end{equation}
On the other hand, it follows from the definition (\ref{Def24ofF}) that 
\begin{equation}
dF=0. \label{EOMF}
\end{equation}

{}For a spacetime with the topology of $\re^n$, the general solution to the equation of motion (\ref{EOMstarF}) is, 
\begin{equation}
 ^*\hspace{-.05cm} F = dB,
 \end{equation}
for some $(D-p-2)$-form $B$.  We will call the original form $A$  the ``electric potential" and the form $B$ just introduced the ``magnetic potential".  

The electric and magnetic potentials are related through the fact that their curvatures are the duals of each other.  One may then rewrite Maxwell's equations in the form,
\begin{equation} 
^*\hspace{-.05cm} F = H, \; \; \; (-1)^{(p+1)(D-1)-1} \, ^*\hspace{-.05cm} H = F,
\end{equation}
where 
\begin{equation}
H = dB \label{HcurvB0}
\end{equation}
is the curvature of $B$. Here, we have used the identity $^{**} \omega = (-1)^{k(D-1) -1}\omega$ where $\omega$ is a $k$-form in a $D$-dimensional Minkowski spacetime.  In matrix form, 
\begin{equation}
{\mathcal F} = {\mathcal S}  \, ^*\hspace{-.05cm}{\mathcal F}, \label{122}
\end{equation}
where, 
\vspace{.5cm}
\begin{equation}
{\mathcal F} = 
 \begin{pmatrix} F\\ H\\ \end{pmatrix}, \; \; \; {\mathcal S}  = \begin{pmatrix} 0&(-1)^{(p+1)(D-1)-1} \\ 1 & 0 \end{pmatrix}  .\label{123}
\end{equation}
\vspace{.3cm}

\noindent
One refers to (\ref{122}) as the twisted self-dual formulation of Maxwell's equations \cite{Cremmer:1998px}.

All the steps and concepts are already contained in the case $p=1$ which we will treat in detail to avoid unnecessary cluttering with indices. We will give at the end of the section the results for the general case.

\subsection{The Case $p=1$} 
When $p = 1$, the action (\ref{ActionpD0}) reduces to
\begin{equation}
S[A_\mu] = - \frac{1}{4} \int d^Dx   F_{\mu \nu} F^{\mu \nu} , \label{Action1D0}
\end{equation}
with 
\begin{equation}
F_{\mu \nu} = \partial_\mu A_\nu - \partial_\nu A_\mu.
\end{equation}

The corresponding Hamiltonian form is,
\begin{equation}
S[A_i, \pi^i, A_0] = \int d^Dx\left(\pi^i \dot{A}_i  - {\mathcal H} - A_0 {\mathcal G} \right) \, , \label{HamAction0} 
\end{equation}
with, 
\begin{equation}
{\mathcal H} = \frac{1}{2} \left( {\mathcal E}^k{\mathcal E}_k + \frac{1}{(D-3)!}{\mathcal B}^{k_1 \cdots k_{D-3}}{\mathcal B}_{k_1 \cdots k_{D-3}} \right) \, ,
\end{equation}
and,
\begin{equation}
{\mathcal G} = - \pi^k_{\; \; ,k}   \; .
\end{equation}
Here, the electric field ${\mathcal E}^k$ is just the conjugate momentum $\pi^k$,
\begin{equation}
{\mathcal E}^k = \pi^k \, , \label{EPi0}
\end{equation}
while the magnetic field ${\mathcal B}^{k_1 \cdots k_{D-3}}$ is given by,
\begin{equation}
{\mathcal B}^{k_1 \cdots k_{D-3}} = \frac{1}{2} \epsilon^{k_1 \cdots k_{D-3}mn}F_{mn} \, .
\end{equation}
When the Hamiltonian equations of motion hold, one finds ${\mathcal E}^k = -F^{0k} $.

The gauge transformations read
\begin{eqnarray}
&&\delta_\Lambda A_i  = \partial_k \Lambda  \label{gauge10}\\
&& \delta_\Lambda \pi^i =  0. \label{gauge20}
\end{eqnarray}

\subsubsection{Magnetic Potential}

The solution of the constraint ${\mathcal G} = 0$ is,
\begin{equation}
\pi^k  = \frac{1}{(D-3)!}\epsilon^{kj_1j_2 \cdots j_{D-2}} \partial_{[j_1} B_{j_2 \cdots j_{D-2}]} \, ,\label{DefPhi0}
\end{equation}
and it brings in a $(D-3)$-form $B_{j_1 \cdots j_{D-3}}$ which is the magnetic dual of $A_i$.  

Since the electric field is gauge invariant, the $(D-3)$-form $B_{j_1 \cdots j_{D-3}}$ may be assumed not to transform under the gauge transformations (\ref{gauge10}) and (\ref{gauge20}).  However, since only the field strength, 
\begin{equation}
H_{j_1j_2 \cdots j_{D-2}} = (D-2) \partial_{[j_1} B_{j_2 \cdots j_{D-2}]}, \label{defHk0}
\end{equation}
of $B_{j_1 \cdots j_{D-3}}$ appears, the expression (\ref{DefPhi0}) is invariant if one transforms $B_{j_1 \cdots j_{D-3}}$ as
\begin{equation}
\delta_{\tilde{\Lambda}} B_{j_1 \cdots j_{D-3}} = (D-3) \partial_{[j_1} \tilde{\Lambda}_{j_2 \cdots j_{D-3}]} \, , \label{GaugePhi0}
\end{equation}
where $\tilde{\Lambda}_{j_1 \cdots j_{D-4}}$ is an arbitrary $(D-4)$-form.
The gauge invariant field strength (\ref{defHk0})
coincides, up to the sign facor $(- 1)^{D-2}$, with the spatial dual of the electric field ${\mathcal E}^k$ of the original one-potential (electric) formulation,
\begin{eqnarray}
&& H_{j_1j_2 \cdots j_{D-2}} = (- 1)^{D-2}\epsilon_{j_1j_2 \cdots j_{D-2}m} {\mathcal E}^m,  \\
&&  {\mathcal E}^k = \frac{1}{(D-2)!}\epsilon^{kj_1j_2 \cdots j_{D-2}} H_{j_1j_2 \cdots j_{D-2}},
\end{eqnarray}
and fulfills,
\begin{equation}
\partial_{[j_1} H_{j_2 \cdots j_{D-1}]} = 0.
\end{equation}

\subsubsection{Two-Potential Action}
We now pass to show how our systematic procedure leads to the two-potential action first postulated in \cite{ScSe} as an extension of the untwisted self-duality action of \cite{Henneaux:1988gg}.

In terms of the electric and magnetic potentials $(A_k, B_{j_1 \cdots j_{D-3}})$, the action (\ref{HamAction0}) takes the form,
\begin{equation}
S[A_k, B_{j_1 \cdots j_{D-3}}] = \int d^Dx \left(\frac{1}{(D-2)!}\epsilon^{kj_1j_2 \cdots j_{D-2}} H_{j_1j_2 \cdots j_{D-2}} \dot{A}_k  - {\mathcal H} \right), \label{HamAction000}
\end{equation}
with,
\begin{equation}
{\mathcal H} = \frac{1}{2} \left(  \frac{1}{(D-2)! }H_{j_1j_2 \cdots j_{D-2}} H^{j_1j_2 \cdots j_{D-2}}+ \frac{1}{(D-3)!}{\mathcal B}^{k_1 \cdots k_{D-3}}{\mathcal B}_{k_1 \cdots k_{D-3}}  \right) \, . \hspace{1.5cm}\label{Energy100}
\end{equation}
One may give a manifestly gauge invariant form to (\ref{HamAction000}),
\begin{equation}
S[A_k, B_{j_1 \cdots j_{D-3}}] = \int d^Dx \left(\frac{1}{(D-2)!}\epsilon^{kj_1j_2 \cdots j_{D-2}} H_{j_1j_2 \cdots j_{D-2}} F_{0k}  - {\mathcal H} \right), \label{SHamActionTer00}
\end{equation}
where,
 \begin{equation}
F_{0k} = \partial_0 A_k - \partial_k A_0. \label{DefF0k00}
\end{equation}
Expressions (\ref{HamAction000}) and (\ref{SHamActionTer00}) coincide because the temporal component $A_0$ appears only through a total derivative.

\subsubsection{Two-Potential Equations of Motion}
The equations of motion that follow from demanding that the action be stationary are
\begin{eqnarray}
&&\partial_k \left(H^{ki_1 \cdots i_{D-3}} + \epsilon^{kmi_1 \cdots i_{D-3}} \dot{A}_m \right) = 0 \label{H10}\\
&& \partial_m\left(F^{mk} + \frac{1}{(D-3)!} \epsilon^{mki_1 \cdots i_{D-3}} \dot{B}_{i_1 \cdots i_{D-3}}\right) = 0 \label{H20}
\end{eqnarray}
Equation (\ref{H10}) implies, 
\begin{equation}
H^{ki_1 \cdots i_{D-3}} + \epsilon^{kmi_1 \cdots i_{D-3}} \dot{A}_m = \epsilon^{kmi_1 \cdots i_{D-3}}  \partial_m A_0,
\end{equation} 
for some function $A_0$, in terms of which, recalling (\ref{DefF0k00}), one can write
\begin{equation}
H^{ki_1 \cdots i_{D-3}} + \epsilon^{kmi_1 \cdots i_{D-3}} F_{0m} = 0. \label{HamilTwisted10}
\end{equation} 
Similarly,  Eq. (\ref{H20}) implies,
\begin{equation}
F^{mk} + \frac{1}{(D-3)!} \epsilon^{mki_1 \cdots i_{D-3}} \dot{B}_{i_1 \cdots i_{D-3}} = \frac{1}{(D-4)!} \epsilon^{mki_1 \cdots i_{D-3}} \partial_{i_1}B_{0i_2 \cdots i_{D-3}}, \label{HamilTwisted20}
\end{equation}
for some functions $B_{0i_2 \cdots i_{D-3}}$.  Defining
\begin{equation}
H_{0i_1i_2 \cdots i_{D-3}} = \dot{B}_{i_1 \cdots i_{D-3}} - (D-3) \partial_{[i_1}B_{0i_2 \cdots i_{D-3}]},
\end{equation}
one can rewrite (\ref{HamilTwisted20}) as,
\begin{equation}
F^{mk} + \frac{1}{(D-3)!} \epsilon^{mki_1 \cdots i_{D-3}} H_{0i_1 \cdots i_{D-3}} =0.
\end{equation}
To derive (\ref{HamilTwisted10}) and (\ref{HamilTwisted20}) from (\ref{H10}) and (\ref{H20}), one must use the fact that  the Betti numbers $b_1$ and $b_{D-3}$ of $\re^{D-1}$ vanish.

Eqs. (\ref{HamilTwisted10}) and (\ref{HamilTwisted20}) are the twisted self-duality equations (\ref{122}).  More precisely, they are the purely spatial components of (\ref{122}), but these are equivalent to the full set (\ref{122}).  Indeed, this set is redundant since half of the equations in (\ref{122}) -- which may be thought of as being the equations with one index equal to zero -- are  consequences of the other half -- which may be thought of as the purely spatial equations. Therefore, we have found an action for the twisted self-duality equations, which may be written in the equivalent forms (\ref{HamAction000}) or (\ref{SHamActionTer00}).

\subsubsection{Symplectic Structure}

The Poisson brackets of the magnetic and electric field strengths that follow from the kinetic term in the action  (\ref{HamAction000}) are,
\begin{eqnarray}
&& [{\mathcal B}^{i_1 \cdots i_{D-3}}(x), {\mathcal B}^{j_1 \cdots j_{D-3}}(y)] =  0,\\
&& [{\mathcal B}^{i_1 \cdots i_{D-3}}(x), H_{j_1 \cdots j_{D-2}}(y)]= (-1)^{D-2} \, (D-2)! \, \delta^{i_1 \cdots i_{D-3} k}_{j_1 \cdots j_{D-2}} \; \delta_{,k}(x,y), \hspace{1cm}\\
&& [H_{i_1 \cdots i_{D-2}}(x), H_{j_1 \cdots j_{D-2}}(y)] = 0, \label{2.40}
\end{eqnarray}
where $ \delta^{i_1 \cdots i_{D-3} k}_{j_1 \cdots j_{D-2}}$ is the Kronecker delta in the space of fully antisymmetric tensors of rank $(D-2)$,
\begin{equation}
\delta^{i_1 \cdots i_{D-2} }_{j_1 \cdots j_{D-2}} =  \delta^{[i_1}_{j_1} \delta^{i_2}_{j_2} \cdots \delta^{i_{D-2}]}_{j_{D-2}} .
\end{equation}
One sees that the electric and magnetic field strengths are canonically conjugate.

There is a way to rewrite the kinetic term in the action (\ref{SHamActionTer00}) that makes the twist matrix ${\cal S}$ appear explicitly and exhibits thereby its connection with the symplectic structure.  We start with the observation that $H \wedge F$ is a total derivative,
\begin{equation}
H \wedge F =  d( B \wedge F).
\end{equation}
Now, the spacetime exterior derivative $d$ can be split as $d = d_S + d_t$, where $d_S$ is the spatial exterior derivative and $d_t = dt \frac{\partial}{\partial t}$ is the exterior derivative in the time direction.  Similarly, any form can be split as $A = A_S + A_t$, where $A_S$ is the purely spatial part of $A$, while $A_t$ is the piece linear in $dt$. Therefore,
\begin{eqnarray}
H \wedge F &=&  H_S  \wedge F_t  + H_t  \wedge F_S  \\
&=& H_S  \wedge F_t + F_S  \wedge H_t,
\end{eqnarray}
 since $F_S$ is a 2-form, and therefore commutes with $H_t$.  Here, $H_S = d_S B_S$, $H_t = d_t B_S + d_S B_t$, and similar formulas hold for $F_S$ and $F_t$ in terms of $A_t$ and $A_S$.
 
The kinetic term in the action (\ref{SHamActionTer00})  can be rewritten, after integration by parts, as,
\begin{eqnarray}
&& \int d^Dx \frac{1}{(D-2)!}\epsilon^{kj_1j_2 \cdots j_{D-2}} H_{j_1j_2 \cdots j_{D-2}} F_{0k}  \nonumber \\
&& = \frac{1}{2} \int d^Dx \epsilon^{kj_1j_2 \cdots j_{D-2}} \left(\frac{1}{(D-2)!} H_{j_1j_2 \cdots j_{D-2}} F_{0k} - \frac{1}{2! (D-3)!}  F_{kj_1} H_{0j_2 \cdots j_{D-2}} \right).\nonumber \\
&& \label{kinetic}
 \end{eqnarray}
In terms of forms, the integrand in (\ref{kinetic}) is,
\begin{equation}
K =  \frac{1}{2} (H_S  \wedge F_t - F_S  \wedge H_t), \label{kinetic0} 
\end{equation} and it is similar in form to the topological invariant $H \wedge F$, but differs from it in the relative sign of the second term.
This sign difference makes (\ref{kinetic0}) not to be a total derivative. Note, however,  that in spite of the sign change, $A_t$ and $B_t$ still enter the kinetic term through a total derivative and drop out from the action because, e.g., $F_S \wedge d_S B_t = d_S (F_S \wedge B_t)$ since $d_S F_S = d_S^2 A_S = 0$. 
 
 Collecting the curvatures as $(F^a) = (F,H)$, one finds that,  
 \begin{equation}
K =  \frac{1}{2} {\mathcal S}_{ab} F^a_S \wedge F^b_t, \label{kinetic00} 
\end{equation}  and therefore, the kinetic term of the action -- and hence the symplectic form -- are intimately connected with the twist matrix.  

We conclude this subsection by pointing out that it follows from the previous discussion that adding an arbitrary symmetric matrix $M_{ab}$ to the antisymmetric twist matrix ${\mathcal S}_{ab}$,
 \begin{equation}
K' =  \frac{1}{2} \left( {\mathcal S}_{ab} + M_{ab} \right) F^a_S \wedge F^b_t, \label{kinetic00'} 
\end{equation} 
changes the action by a total derivative.  As we shall see below, it turns out that in the presence of couplings, non-vanishing choices of $M_{ab}$ might be convenient to exhibit explicitly the gauge symmetries.

\subsubsection{Lorentz Invariance and Coupling to Gravity}

The Gauss constraint is not changed by the coupling to gravity because the gauge transformation of a form does not depend on the metric.  One can therefore introduce the magnetic potential in exactly the same way.

The linear momentum (generator of spatial Lie derivatives) obtained from the action of the two-potential theory is,
\begin{equation}
{\mathcal H}_i =  F_{ik}  {\mathcal E}^k = - {\mathcal B}^{j_1 \cdots j_{D-3}} H_{ij_1 \cdots j_{D-3}}.
\end{equation}

The coupling to gravity is achieved by changing the Hamiltonian density ${\mathcal H}$ in the action (\ref{HamAction000}) by, 
\begin{equation}
N^\perp {\mathcal H}_\perp + N^k {\mathcal H}_k,
\end{equation}
where $N^\perp$ and $N^k$ are the lapse and the shift appearing in the Hamiltonian formulation in curved space, and where ${\mathcal H}_\perp$ is given by,
\begin{equation}
{\mathcal H}_\perp = \frac{1}{2} \left(  \frac{1}{(D-2)! }g^{\frac{1}{2}} H_{j_1j_2 \cdots j_{D-2}} H^{j_1j_2 \cdots j_{D-2}}+ \frac{1}{(D-3)!}g^{-\frac{1}{2}}{\mathcal B}^{k_1 \cdots k_{D-3}}{\mathcal B}_{k_1 \cdots k_{D-3}}  \right)  \label{Energy20}
\end{equation}
where the indices are raised or lowered with $g^{ij}$ or $g_{ij}$, respectively. The generators ${\mathcal H}_\perp$ and ${\mathcal H}_i$ obey the algebra,
\begin{eqnarray}
&& [{\mathcal H}(x), {\mathcal H}(y)] = \left({\mathcal H}^i(x) + {\mathcal H}^i(y)\right) \delta_{,i}(x,y), \label{algebra1}\\
&& [{\mathcal H}(x), {\mathcal H}_i(y)] = {\mathcal H}(y)\delta_{,i}(x,y),  \label{algebra2}\\\
&& [{\mathcal H}_i(x), {\mathcal H}_j(y)] ={\mathcal H}_i(y)\delta_{,j}(x,y) + {\mathcal H}_j(x)\delta_{,i}(x,y), \label{algebra3}\
\end{eqnarray}
which shows that the coupling to gravity is generally covariant and, in particular,  that in flat space the theory is Lorentz invariant.  Note that in comparison with the standard Hamiltonian formulation in the electric representation, there is no Gauss constraint in the right-hand side of the algebra since here Gauss' law is identically satisfied.

\subsection{The Case $0 < p <D-2$}
We now pass to show how, also in this case, our systematic procedure leads to the two-potential action first postulated in \cite{ScSe} as an extension of the untwisted self-duality action of \cite{Henneaux:1988gg}. 

By following the same steps as in the case $p=1$, one obtains the two-potential action 
\begin{equation}
S[A_{k_1 \cdots k_p}, B_{j_1 \cdots j_{D-p-2}}] = \int d^Dx \left(\frac{\epsilon^{k_1\cdots k_p j_1 \cdots j_{D-p-1}}}{p! \,(D-p-1)!} H_{j_1 \cdots j_{D-p-1}} \dot{A}_{k_1 \cdots k_p}  - {\mathcal H} \right), \label{HamAction0001}
\end{equation}
with,
\begin{equation}
{\mathcal H} = \frac{1}{2} \left( \frac{1}{(D-p-1)!} H_{j_1 \cdots j_{D-p-1}} H{j_1 \cdots j_{D-p-1}} + \frac{1}{(D-p-2)!} {\mathcal B}^{j_1 \cdots j_{D-p-2}} {\mathcal B}_{j_1 \cdots j_{D-p-2}}\right) \, . \label{Energy101}
\end{equation}
Here, $H_{j_1 \cdots j_{D-p-1}}$ is the gauge invariant field strength of the magnetic potential,
\begin{equation}
H_{j_1 j_2 \cdots j_{D-p-1}}= (D-p-1) \partial_{[j_1} B_{j_2 \cdots j_{D-p-1}]}
\end{equation}
(equal on-shell to $\pm$ the spatial dual of $F_{0 k_1 \cdots k_{p-1}}$), while ${\mathcal B}^{j_1 \cdots j_{D-p-2}}$ is the magnetic field,
\begin{equation}
{\mathcal B}^{j_1 \cdots j_{D-p-2}} = \frac{1}{(p+1)!} \epsilon^{j_1\cdots j_{D-p-2} k_1 \cdots k_{p+1}} F_{k_1 \cdots k_{p+1}}
\end{equation}

Again, one may give a manifestly gauge invariant form to (\ref{HamAction0001}),
\begin{equation}
S[A_{k_1 \cdots k_p}, B_{j_1 \cdots j_{D-p-2}}] = \int d^Dx \left(\frac{\epsilon^{k_1\cdots k_p j_1 \cdots j_{D-p-1}}}{p! \,(D-p-1)!} H_{j_1 \cdots j_{D-p-1}} F_{0k_1 \cdots k_p}  - {\mathcal H} \right), \label{SHamActionTer001}
\end{equation}
since in this expression, the temporal component $A_{0k_1 \cdots kÑ{p-1}}$ appears only through a total derivative.

All the comments and conclusions of the previous subsection go through unchanged.  In particular, the fact that the integrand of the kinetic term can be written as,
 \begin{equation}
K =  \frac{1}{2} {\mathcal S}_{ab} F^a_S F^b_t \label{kinetic00p} 
\end{equation}
(up to a total derivative), where ${\mathcal S}_{ab}$ is the (antisymmetric or symmetric) ``twisting" matrix appearing in the twisted self-duality equations, remains true.  

\subsection{The Case $p=0$ or $p=D-2$}
In the case $0<p<D-2$, one could have started from the electric formulation and introduce the magnetic potential by solving the Gauss electric constraint, or conversely, one could have started from the magnetic formulation, solve the magnetic Gauss constraint and introduce the electric potential.  However, when $p=0$, there is no constraint to be solved in the electric formulation and when $p=D-2$, there is no constraint to be solved in the magnetic formulation.  

Nevertheless, one can fit these ``marginal cases" in the present treatment by slightly streching the argument.  One cannot take over the form of the constraint equations from the generic dimensions because, as we just said, those equations are not present. However, one can take over the form of their solutions. That is, if we start from the electric formulation, we set, when $p=0$,
\begin{equation}
\pi_A = \partial_j\left(\epsilon^{ji_1i_2 \cdots i_{D-2}} B_{i_1i_2 \cdots i_{D-2}} \right), \label{piScalar}
\end{equation}
in order to introduce the magnetic potential, which can always be done since it does not restrict $\pi_A$. The resulting key formulas of the previous subsection hold then unchanged.  Conversely, if one had started from the magnetic formulation for $p=D-2$, the magnetic momentum would be a scalar density and one would write 
\begin{equation}
\pi_B = \partial_j\left(\epsilon^{ji_1i_2 \cdots i_{D-2}} A_{i_1i_2 \cdots i_{D-2}} \right), \label{piScalar'}
\end{equation}
 to introduce the electric potential $A_{i_1i_2 \cdots i_{D-2}}$.

\subsection{The cases $p=D-1$ and $p \geq D$}
The cases $p=D-1$ and $p \geq D$ do not fit in the present treatment. When $p=D-1$, the constraints imply that there are no local degrees of freedom.   When $p=D$, the curvature is identically zero and so is the action.  There are again no local degrees of freedom. Both cases belong with the topological considerations of \cite{BHtoappear}. When $p > D$, the problem is empty because $A \equiv 0$.

\section{Introduction of a Chern-Simons Term}
\setcounter{equation}{0}

This section is devoted to the inclusion of a Chern-Simons term. We will again analyze in detail the simplest case, that is, $p=1$, $D=3$, and then indicate explicitly the results for the generalization to  $p=3$, $D=11$, which is of special interest because it arises in supergravity. The procedure applies however to all the other cases. 
 
For the case of a $3$-form with Chern-Simons couplings in eleven dimensions, our action coincides with the one given earlier in \cite{Bandos} when the additional gauge freedom is fixed in a very simple form.

\subsection{The Simplest Setting: Maxwell-Chern-Simons Action in 3 Dimensions}
\label{MCS3D}

It turns out that, as it is often the case, many of the key aspects are present in the simplest low dimensional model:  This subsection is devoted to analyze the problem in three-dimensional spacetime.   

The twisted self-duality equations take the form (\ref{122}) with the definition (\ref{HcurvB0}) modified to read \cite{Cremmer:1998px},
\begin{equation}
H = dB - 4 \alpha A . \label{HcurvB1}
\end{equation}

\subsubsection{One-Potential Action}

The Lagrangian form of the Maxwell-Chern-Simons action is \cite{Deser:1982vy},
\begin{equation}
S[A_\mu] = \int d^3x\left( - \frac{1}{4}  F_{\mu \nu} F^{\mu \nu} -\alpha \epsilon^{\lambda \mu \nu} F_{\lambda \mu} A_\nu \right) \, ,
\end{equation}
and the corresponding Hamiltonian form is,
\begin{equation}
S[A_i, \pi^i, A_0] = \int d^3x\left(\pi^i \dot{A}_i  - {\mathcal H} - A_0 {\mathcal G} \right) \, , \label{HamAction} 
\end{equation}
with, 
\begin{equation}
{\mathcal H} = \frac{1}{2} \left( {\mathcal E}^k{\mathcal E}_k + {\mathcal B}^2 \right) \, ,
\end{equation}
and,
\begin{equation}
{\mathcal G} = - \pi^k_{\; \; ,k} - \alpha \, \epsilon^{km} F_{km} = - \left( \pi^k + 2\alpha \epsilon^{km} A_m\right)_{,k} \; . \label{constraintMCS3}
\end{equation}
Here, the electric field ${\mathcal E}^k$ is related to the conjugate momentum $\pi^k$ through
\begin{equation}
{\mathcal E}^k = \pi^k - 2\alpha \epsilon^{km} A_m \, , \label{EPi}
\end{equation}
while the magnetic field ${\mathcal B}$ is given by,
\begin{equation}
{\mathcal B} = \frac{1}{2} \epsilon^{mn}F_{mn} \, .
\end{equation}
We use the convention $\epsilon_{012} = 1 = - \epsilon^{012}$. One has ${\mathcal E}^k = -F^{0k} $ when the Hamiltonian equations of motion hold.

We see from (\ref{constraintMCS3}) that the gauge generator ${\mathcal G}$ remains the divergence of a local vector density, as required by the duality principle when implemented according to our procedure. 

The gauge transformations read
\begin{eqnarray}
&&\delta_\Lambda A_i  = \partial_k \Lambda  \label{gauge1}\\
&& \delta_\Lambda \pi^i =  2 \alpha \epsilon^{km} \partial_m \Lambda \label{gauge2}
\end{eqnarray}
Contrary to what happens in the case with no Chern-Simons term, the conjugate momentum $\pi^i$ is no longer gauge invariant.  But the electric field ${\mathcal E}^i$ remains so. 

\subsubsection{Magnetic Potential}

The solution of the constraint ${\mathcal G} = 0$ is,
\begin{equation}
\pi^k + 2 \alpha \epsilon^{km} A_m = \epsilon^{km} \partial_m B \, ,\label{DefPhi}
\end{equation}
and it brings in a scalar field $B$, which is the magnetic dual of $A_i$.  

The gauge transformation for $B$ will be taken to be, 
\begin{equation}
\delta_\Lambda B = 4 \alpha \Lambda \, , \label{GaugePhi}
\end{equation}
which solves the variation of (\ref{DefPhi}) given (\ref{gauge1}) and (\ref{gauge2}).  For an open space, this equation incorporates the requirement that the gauge transformation should be ``proper" in the sense of \cite{Regge:1974zd,Benguria:1976in}.  For a compact space, other additional considerations are needed, which will be adressed in \cite{BHtoappear}.
The gauge invariant field strength of the magnetic potential $B$ is,
\begin{equation}
H_k = \partial_k B - 4 \alpha A_k, \label{defHk}
\end{equation}
and coincides, through (\ref{EPi}) and (\ref{DefPhi}), with the negative of the spatial dual of the electric field $E_k$ of the original one-potential (electric) formulation,
\begin{equation}
H_k = - \epsilon_{km} {\mathcal E}^m,  \; \; \; \; \; {\mathcal E}^k = \epsilon^{km} H_m.
\end{equation}
It follows from its definition (\ref{defHk}) that the gauge invariant field strength $H_k$ fulfills,
\begin{equation}
\partial_i H_j - \partial_j H_i = - 4 \alpha F_{ij} = - 4 \alpha \epsilon_{ij} {\mathcal B}
\end{equation}

\subsubsection{Two-Potential Action}

We now proceed to obtain the two-potential action.

In terms of the electric and magnetic potentials $(A_k, B)$, the action (\ref{HamAction}) takes the form,
\begin{equation}
S[A_k, B] = \int d^3x \left(\epsilon^{km} \partial_m B \dot{A}_k - 2 \alpha \epsilon^{km} A_m \dot{A}_k - {\mathcal H} \right), \label{HamAction00}
\end{equation}
with,
\begin{equation}
{\mathcal H} = \frac{1}{2} \left( H^kH_k + {\mathcal B}^2 \right) \, . \label{Energy1}
\end{equation}
Through integration by parts, one may rewrite (\ref{HamAction00}) as,
\begin{equation}
S[A_k, \phi] = \int d^3x \left(\frac{1}{2} \epsilon^{km} H_m \dot{A}_k - \frac{1}{4} \epsilon^{km} F_{km} \dot{B} - {\mathcal H} \right), \label{SHamActionBis}
\end{equation}
an expression in which only the gauge invariant field strengths and the time derivatives of $A_k$ and $B$ appear.
One may give a manifestly gauge invariant form to (\ref{SHamActionBis}),
\begin{equation}
S[A_\mu, \phi] = \int d^3x \left(\frac{1}{2} \epsilon^{km} H_m F_{0k} - \frac{1}{4} \epsilon^{km} F_{km} H_0 - {\mathcal H} \right), \label{SHamActionTer}
\end{equation}
where,
 \begin{eqnarray}
&& H_0 = \partial_0 B - 4 \alpha A_0, \label{DefH0}\\
&& F_{0k} = \partial_0 A_k - \partial_k A_0. \label{DefF0k}
\end{eqnarray}
Expressions (\ref{SHamActionBis}) and (\ref{SHamActionTer}) coincide because the temporal component $A_0$ appears only through a total derivative.  Note again the emergence of the structure $\frac{1}{2} {\mathcal S}_{ab} F^a_S F^b_t$, where the curvatures are now the full gauge invariant curvatures.

\subsubsection{Two-Potential Equations of Motion}
The equations of motion that follow from demanding that the action be stationary are
\begin{eqnarray}
&&\partial_k \left(H^k + \epsilon^{km} \dot{A}_m \right) = 0 \label{H1}\\
&&  - \epsilon^{km} \partial_m\left(\dot{B} +  {\mathcal B}\right) + 4\alpha \left(H^k + \epsilon^{km} \dot{A}_m \right)= 0 \label{H2}
\end{eqnarray}
Equation (\ref{H1}) implies, 
\begin{equation}
H^k + \epsilon^{km} \dot{A}_m = \epsilon^{km} \partial_m A_0,
\end{equation} 
for some function $A_0$, in terms of which, recalling (\ref{DefH0}) and (\ref{DefF0k}), one can therefore write
\begin{equation}
H^k + \epsilon^{km} F_{0m} = 0. \label{HamilTwisted1}
\end{equation} 
Taking (\ref{HamilTwisted1}) into account,  Eq. (\ref{H2}) becomes,
\begin{equation}
\partial_m \left(H_0 + {\mathcal B} \right) = 0,
\end{equation}
which implies,
\begin{equation}
H_0 + {\mathcal B} = 0. \label{HamilTwisted2}
\end{equation}
Again, just as when we established (\ref{GaugePhi}), one must impose boundary conditions at infinity or make additional special considerations for compact spaces \cite{BHtoappear}.

Eqs. (\ref{HamilTwisted1}) and (\ref{HamilTwisted2}) are the twisted self-duality equations (\ref{122}) with $H$ given by (\ref{HcurvB1}).  Therefore, we have found an action for them, which may be written in the equivalent forms (\ref{HamAction00}), (\ref{SHamActionBis}) or (\ref{SHamActionTer}).

\subsubsection{Lorentz Invariance and Coupling to Gravity}
The Poisson brackets of the electric and magnetic field strengths that follow from the action  (\ref{HamAction00}) are
\begin{eqnarray}
&& [{\mathcal B}(x), {\mathcal B}(y)] =  0,\\
&& [{\mathcal B}(x), H_k(y)]= - \delta_{,k}(x,y) \\
&& [H_k(x), H_m(y)] = - 4 \alpha \epsilon_{km} \delta(x,y) \label{3.29}
\end{eqnarray}
Comparing (\ref{3.29}) with (\ref{2.40}), we see that when $\alpha \not= 0$, the magnetic strengths have non zero bracket among themselves.  Therefore, a purely magnetic representation of the Maxwell-Chern-Simons theory does not exist.

The linear momentum (generator of spatial Lie derivatives) obtained from the action of the two-potential theory is,
\begin{equation}
{\mathcal H}_i = \epsilon^{km} F_{ik}  H_m = - {\mathcal B} H_i.
\end{equation}

The coupling to gravity is achieved by changing the Hamiltonian density ${\mathcal H}$ in the action (\ref{HamAction00}) by, 
\begin{equation}
N^\perp {\mathcal H}_\perp + N^k {\mathcal H}_k,
\end{equation}
where $N^\perp$ and $N^k$ are the lapse and the shift appearing in the Hamiltonian formulation in curved space, and where ${\mathcal H}_\perp$ is given by,
\begin{equation}
{\mathcal H}_\perp = \frac{1}{2} \left(g^{\frac{1}{2}} g^{ij} H_iH_j + g^{-\frac{1}{2}}{\mathcal B}^2 \right).  \label{Energy2}
\end{equation}
The generators ${\mathcal H}_\perp$ and ${\mathcal H}_i$ obey the algebra (\ref{algebra1}), (\ref{algebra2}) and (\ref{algebra3}), which shows that the coupling to gravity is generally covariant and, in particular,  that in flat space the theory is Lorentz invariant.

\subsubsection{External $p$-Form Field}
\label{discr}

The authors of \cite{ScSe} considered an extension of the free theory in which the field strengths are modified by the addition of a ``Chern-Simons"-like form $\Omega$. If one were to take this form $\Omega$ as a prescribed external field, then the corresponding equation of motion would be Eq. (2.38)  of \cite{ScSe}, which is indeed a twisted self-duality condition.  However, this external field setting (which the present method could also handle) is quite different from the Maxwell-Chern-Simons theory considered here, which is a closed system. One might nevertherless wonder whether a blind application of the formulas of \cite{ScSe} to the standard Maxwell-Chern-Simons theory leads to the correct action.  It turns out that this is not the case.   

This can be seen as follows. The action of \cite{ScSe}  is written in terms of the $(p+1)$-form $\Omega$ appearing in the modified field strengths, which, for the simplest case treated in this subsection, is a $2$-form that one obtains from (\ref{HcurvB1}) to be,
\begin{equation}
\Omega = - 4 \, \alpha \; ^*\hspace{-.05cm} A\, . \label{Omega}
\end{equation}
Inserting the expression (\ref{Omega}) into the integral (2.39) of \cite{ScSe}, one finds,
\begin{equation}
2 \, \alpha \int d^3x \left(\epsilon^{ij} F_{ij} A_0 - 2 A^k \partial_k B\right) \label{SSInt}
\end{equation}
The integral (\ref{SSInt}) is to be compared with the difference between (\ref{HamAction00}) and the free action.  There are several key differences that prevent one from reconciling both expressions, namely: (i) The integral (\ref{SSInt}) depends on $A_0$ and therefore it is not gauge invariant.  In contradistinction, the counterpart to (\ref{SSInt}) in our action does not depend on $A_0$, and it is gauge invariant; (ii) Even in the $A_0 = 0$ gauge, the functional forms are essentially different.  For example, (\ref{SSInt}) is only linear in $\alpha$, whereas our action contains as well a piece proportional to $\alpha^2$.

We see therefore no escape to the conclusion that (\ref{SSInt}) does not lead to the two-potential version of the standard Chern-Simons action. On the other hand, the action derived in the previous subsubsections by our systematic procedure, does.  This analysis goes through unchanged in the more general cases discussed below.

\subsection{Maxwell-Chern-Simons Action For a $3$-Form in Eleven Dimensions}

\subsubsection{One-Potential Action}

The standard single-potential Maxwell-Chern-Simons action is given by,
\begin{equation}
S[A_{\lambda \mu \nu}] = \int d^{11}x\left( - \frac{1}{2 \cdot 4!}  F_{\lambda \mu \nu \rho} F^{\lambda \mu \nu \rho} -\alpha \epsilon^{\lambda_1 \lambda_2 \cdots \lambda_{11}} F_{\lambda_1 \cdots \lambda_4} F_{\lambda_5 \cdots \lambda_8}A_{\lambda_9 \lambda_{10}\lambda_{11}}\right) \, ,
\end{equation}
with, 
\begin{equation}
F_{\lambda \mu \nu \rho} = 4 \partial_{[\lambda} A_{\mu \nu \rho]}.
\end{equation}
The square bracket indicates complete antisymmetrization in the enclosed indices normalized by dividing by the appropriate factorial so that it is idempotent. We set $ \epsilon_{0 \, 1 \, \cdots \, 9 \, 10} = 1 =  - \epsilon^{0 \, 1 \, \cdots \, 9 \, 10}$.

The twisted self-duality equations take again the form (\ref{122}) with the definition (\ref{HcurvB0}) modified to read \cite{Cremmer:1998px},
\begin{equation}
H=  dB - 3 \left(3! (4!)^2\right)  \alpha A \wedge F. 
\end{equation}

The Hamiltonian action is
\begin{equation}
S[A_{ijk}, \pi^{ijk}, A_{0ij}] = \int d^{11}x\left( \pi^{ijk} \dot{A}_{ijk}  - {\mathcal H} - \frac{1}{2!}A_{0ij} {\mathcal G}^{ij} \right) \, , \label{HamAction3} 
\end{equation}
with 
\begin{equation}
{\mathcal H} = \frac{1}{2} \left( \frac{1}{3!} {\mathcal E}^{ijk}{\mathcal E}_{ijk} +\frac{1}{6!} {\mathcal B}^{i_1 \cdots i_6} {\mathcal B}_{i_1 \cdots i_6} \right) \, ,
\end{equation}
and
\begin{equation}
{\mathcal G}^{ij} = - 6\pi^{kij}_{\; \; \; \; \; \; ,k} - 6 \alpha \, \epsilon^{ijk_1 \cdots k_8} F_{k_1 \cdots k_4} F_{k_5 \cdots k_8}= -6 \left( \pi^{kij} +4\alpha \epsilon^{kijm_1 \cdots m_7} F_{m_1 \cdots m_4} A_{m_5 m_6 m_7}\right)_{,k} \; .
\end{equation}
Here, the electric field ${\mathcal E}^{ijk}$ is related to the conjugate momentum $\pi^{ijk}$ through
\begin{equation}
\frac{1}{3!}{\mathcal E}^{ijk} = \pi^{ijk} - 8\alpha \epsilon^{ijki_1\cdots i_7} F_{i_1 \cdots i_4} A_{i_5 i_6 i_7} \, , \label{EPi3}
\end{equation}
while the magnetic field $B^{i_1 \cdots i_6}$ is given by
\begin{equation}
B^{i_1 \cdots i_6} = \frac{1}{4!} \epsilon^{i_1 \cdots i_6 j_1 j_2 j_3 j_4}F_{j_1 j_2 j_3 j_4} \, .
\end{equation}
One has ${\mathcal E}^{ijk} = -F^{0ijk} $ on Hamiltonian shell.

The gauge transformations read
\begin{eqnarray}
&&\delta_\Lambda A_{ijk}  = 3 \partial_{[i} \Lambda_{jk]}  \label{gauge13}\\
&& \delta_\Lambda \pi^{ijk} =  24 \alpha \epsilon^{ijkm_1 \cdots m_7} F_{m_1 \cdots m_4} \partial_{[m_5} \Lambda_{m_6 m_7]} \label{gauge23}
\end{eqnarray}

\subsubsection{Magnetic Potential}
The solution of the constraint ${\mathcal G}^{ij} = 0$ is,
\begin{equation}
\pi^{ijk} +4\alpha \epsilon^{ijkm_1 \cdots m_7} F_{m_1 \cdots m_4} A_{m_5 m_6 m_7} =  \frac{1}{3!\, 6!} \epsilon^{ijkm_1 \cdots m_7}\partial_{m_1}B_{m_2 \cdots m_7} \, ,\label{DefPhi3}
\end{equation}
and it brings in a $6$-form $B_{i_1 \cdots i_6}$ which is the magnetic dual of $A_{ijk}$.  

The gauge transformation for $B_{i_1 \cdots i_6}$ will be taken to be, 
\begin{equation}
\delta_{\Lambda, \tilde{\Lambda}} B_{i_1 \cdots i_6} = 6 \left(\partial_{[i_1} \tilde{\Lambda}_{i_2  \cdots i_6]} + 3! \, 6! \, 6  \alpha F_{[i_1 \cdots i_4}\Lambda_{i_5 i_6]} \right) \, . \label{GaugePhi3}
\end{equation}

The gauge invariant field strength of the magnetic potential $B_{i_1 \cdots i_6}$ is
\begin{equation}
H_{i_1  \cdots i_7} = 7 \left( \partial_{[i_1} B_{i_2 \cdots i_7]} - 12 \alpha \, 3! \, 6! F_{[i_1 \cdots i_4}A_{i_5 i_6 i_7]}\right) \label{DefH7}
\end{equation}
and coincides through (\ref{EPi3}) and (\ref{DefPhi3}) with the negative of the spatial dual of the electric field ${\mathcal E}^{ijk}$ of the original one-potential (electric) formulation. One gets from the definition (\ref{DefH7})
\begin{equation}
\partial_{[i_0}H_{i_1  \cdots i_7]} = - 3 \alpha \, 3! \, 7! F_{[i_1 i_2 i_3 i_4}F_{i_0i_5 i_6 i_7]}
\end{equation}

\subsubsection{Two-Potential Action}
In terms of the electric and magnetic potentials $(A_{k_1 k_2 k_3}, B_{i_1 \cdots i_6})$, the action (\ref{HamAction3}) takes the form,
\begin{eqnarray}
&& S[A_{k_1 k_2 k_3}, B_{i_1 i_2 i_3 i_4 i_5 i_6}] = \nonumber \\
&&  \int d^{11}x  \left(\frac{1}{3!\, 6!} \epsilon^{ijkm_1 \cdots m_7}\partial_{m_1}B_{m_2 \cdots m_7} - 4\alpha \epsilon^{ijkm_1 \cdots m_7} F_{m_1 \cdots m_4} A_{m_5 m_6 m_7} \right)\dot{A}_{ijk} \nonumber \\
&& \hspace{4cm} - \int d^{11}x {\mathcal H} , \label{HamAction003}
\end{eqnarray}
with,
\begin{equation}
{\mathcal H} = \frac{1}{2} \left(\frac{1}{7!} H_{i_1 \cdots i_7}H^{i_1 \cdots i_7} + \frac{1}{6!}{\mathcal B}^{i_1 \cdots i_6} {\mathcal B}_{i_1 \cdots i_6}\right) \, .
\end{equation}

As in the 3-dimensional case, one may give a manifestly gauge invariant form to (\ref{HamAction003}).  Using form notations to avoid lengthy formulas, one finds,
\begin{eqnarray}
&& S[A_{\mu_1 \mu_2 \mu_3}, B_{\mu_1 \mu_2 \mu_3 \mu_4 \mu_5 \mu_6}] = \nonumber \\
&& \; \frac{1}{2} \int \left( H_S \wedge F_t - F_S \wedge H_t + \frac{1}{3} (H_S \wedge F_t + F_S \wedge H_t) \right) \nonumber \\
&& \hspace{4cm} - \int d^{11}x {\mathcal H} \label{SHamActionTer3}
\end{eqnarray}
where the temporal components of the curvatures are,
 \begin{eqnarray}
&& H_{0m_1 \cdots m_6} = \partial_0 B_{m_1 \cdots m_6} + 6 \partial_{[m_1} B_{m_2 \cdots m_6 ]0} \nonumber \\
&& \hspace{1.2cm} - 12 \alpha \, 3! \, 6! \left( 4 F_{0 [m_1 m_2 m_3} A_{m_4 m_5 m_6]}+ 3 F_{[m_1 \cdots m_4} A_{m_5 m_6]0}\right), \label{DefH03}\\
&& F_{0i_1i_2i_3} = \partial_0 A_{i_1 i_2 i_3} - 3\partial_{[i_1} A_{i_2 i_3 ]0}. \label{DefF0k3}
\end{eqnarray}
The two expressions (\ref{HamAction003}) and (\ref{SHamActionTer3}) coincide because the temporal components of the electric and magnetic potentials drop out (they appear only through a total derivative).

The Poisson brackets of the electric and magnetic field strengths that follow from the action  (\ref{HamAction003}) are
\begin{eqnarray}
&& [{\mathcal B}^{i_1\cdots i_6}(x), {\mathcal B}^{j_1 \cdots j_6}(y)] =  0,\\
&& [{\mathcal B}^{i_1 \cdots i_6}(x), H_{k_1 \cdots k_7}(y)]= 7! \,  \delta^{i_1 \cdots i_6 i_7}_{\; k_1 \cdots k_7}\delta_{,i_7}(x,y) \\
&& [H_{k_1 \cdots k_7}(x), H_{m_1 \cdots m_7}(y)]   \nonumber \\
&& \hspace{1.5cm} = -16\alpha \epsilon_{k_1 \cdots k_7 i_1i_2i_3} \epsilon_{m_1 \cdots m_7 j_1j_2j_3}{\mathcal B}^{i_1i_2 i_3 j_1 j_2 j_3}\delta(x,y)
\end{eqnarray}

One easily verifies as in the previous subsection that the variational equations are the twisted self-duality equations.  Therefore, we have found an action for them, which may be written in the equivalent forms (\ref{HamAction003}) or (\ref{SHamActionTer3}).  Similarly, coupling to gravity and demonstration of Lorentz invariance proceed along the same lines. 

The two-potential action discussed in this section is different from that of \cite{ScSe}  in which one would replace the external form $\Omega$ that appears there by the function of the dynamical fields relevant to the case considered here (see subsubsection \ref{discr}).  On the other hand, as it was anticipated in the introduction, the present two-potential action coincides with that of \cite{Bandos}, when the auxiliary vector $v_n$ appearing therein is gauge fixed to have only a non-zero constant time component, i.e., $v_n=(1,0,...,0)$.

\section{Coupled Forms of Different Rank}
\label{FieldStrengthDeformation}
\setcounter{equation}{0}

In this section we show that our procedure can be applied to the coupling among a $1$-form and a $2$-form,   which arises in ten-dimensional Einstein-Maxwell supergravity \cite{Bergshoeff:1981um} ($N=1$, $D=10$ supergravity coupled to  one Maxwell multiplet), and indicate its generalization to couplings of the same type between several $p$-forms of different rank. The dimensional reduction of this case to $4$ dimensions was considered in \cite{ScSe}. We also explain how the procedure can be applied straightforwardly to Pauli couplings to spinors and to couplings to uncharged scalars.

In ten-dimensional Einstein-Maxwell supergravity, one has a $1$-form $A^{(1)}$and a $2$-form $A^{(2)}$ and the part of the action relevant to our problem is,
\begin{equation}
S =- \frac{1}{2} \int d^{10}x \left( \frac{1}{2!} F^{(1)}_{\mu \nu} F^{(1) \mu \nu} +  \frac{1}{3!} F^{(2)}_{\mu \nu \rho} F^{(2) \mu \nu \rho}\right) \label{LEMS}
\end{equation}
where the curvatures are,
\begin{eqnarray}
&& F^{(1)} = dA^{(1)}\\
&& F^{(2)} = dA^{(2)} - \alpha F^{(1)} \wedge A^{(1)}.
\end{eqnarray}
The gauge transformations,  which leave the curvatures invariant, are,
\begin{equation}
\delta_{\Lambda^{(1)},\Lambda^{(2)}} A^{(1)} = d \Lambda^{(1)} , \; \; \; \; \delta_{\Lambda^{(1)},\Lambda^{(2)}} A^{(2)} = d \Lambda^{(2)} + \alpha A^{(1)} \wedge d\Lambda^{(1)}, \label{GaugeSuper}
\end{equation}
where $\Lambda^{(1)}$ and $\Lambda^{(2)}$ are a $0$-form and a $1$-form, respectively.

If one passes to the Hamiltonian form, one obtains the Gauss constraints
\begin{eqnarray}
&& {\mathcal G}_{(1)} = - \partial_j \left( \pi_{(1)}^{j} + 2 \alpha \pi_{(2)}^{ij} A^{(1)}_i \right) \\
&& {\mathcal G}_{(2)}^i = - 2 \partial_j \pi_{(2)}^{ij}.
\end{eqnarray}
where $\pi_{(1)}^{j}$ and $\pi_{(2)}^{ij}$ are the canonical conjugates to $A^{(1)}_i$ and  $A^{(2)}_{ij}$, respectively.  The constraints generate the gauge transformations (\ref{GaugeSuper}).

Both ${\mathcal G}_{(1)}$ and ${\mathcal G}_{(2)}^i$ are local divergences and therefore our procedure can be applied. The magnetic potentials are introduced by solving the Gauss constraints in the form,
\begin{eqnarray}
&& \pi_{(1)}^i  = \frac{1}{7!}\epsilon^{ij_1j_2 \cdots j_{8}} \partial_{[j_1} B_{j_2 \cdots j_{8}]}^{(1)} -  \frac{2 \alpha }{2!\, 6!} \epsilon^{ijm_1 \cdots m_7}\partial_{[m_1}B_{m_2 \cdots m_7}^{(2)}A^{(1)}_{i\, ]}, \label{B7}\\
&& \pi_{(2)}^{ij}  =  \frac{1}{2!\, 6!} \epsilon^{ijm_1 \cdots m_7}\partial_{[m_1}B_{m_2 \cdots m_7]}^{(2)} \, .\label{B6}
\end{eqnarray}
Here $B^{(1)}$ and $B^{(2)}$ are the dual magnetic $7$-form and $6$-form, respectively.

The electric-magnetic action that incorporates the duality principle is again simply the Hamiltonian action  written down  explicitly in \cite{Baulieu:1986hp},  in which one has expressed the conjugate momenta in terms of the magnetic potentials. The equations of motion obtained from the action are the twisted self-duality equations in Hamiltonian form. 

The complete Lagrangian of  ten-dimensional Einstein-Maxwell supergravity differs from the integrand of (\ref{LEMS}) by terms in which the curvatures of $A^{(1)}$ and $A^{(2)}$ are coupled to spinor and scalar fields.  These fields are invariant under the gauge transformations of the $1$-form and the $2$-form.  Therefore the gauge constraints for the complete theory are just those written above and thus the electric-magnetic action action can be completed to the full theory -- a step that will not be taken explicitly in the present work. 

Although it will not be discussed here, the procedure goes through for more complicated supergravities, where interactions of the same type among a collection of $p$-forms appear.  In that case, for the procedure to work, it must be possible to define the gauge transformations for the $p$-forms so that the gauge parameters appear always differentiated.  This requirement is equivalent to demanding that the constraints can be chosen to be local divergences.  It can be shown, following the lines of \cite{Cremmer:1998px}, that this can indeed always be arranged.   For the case of type IIB supergravity, the two-potential action has been discussed in the manifestly Lorentz invariant formalism in \cite{Dall}, where it has been shown explicitly that the equations of motion are the desired ones.  Since, by construction, the same holds true if one applies our method, we conclude that the two actions should  coincide when the auxiliary gauge freedom of the manifestly Lorentz invariant formalism is appropriately fixed.

 Finally we would like to emphasize that  for Yang-Mills  couplings, the procedure does not go through because, in the gauge transformations, the gauge parameter appears undifferentiated.

\section{Conclusion}
\label{Conclusions}

This paper has been devoted to providing a systematic derivation from the Maxwell action of the action principle which yields the condition of electric-magnetic self-duality as its equation of motion.  It is hoped that our results will help dispel the widespread misconception that twisted (and untwisted) self-duality can only be discussed at the level of the equations of motion. 

In the pure Maxwell case we recover in this way an action that had been  postulated by other authors  \cite{ScSe} by boldly extending the one given earlier by us \cite{Henneaux:1988gg} for untwisted self-duality. However, when standard Chern-Simons couplings are brought in - a case that \cite{ScSe} does not claim to describe --, the action we derive is new.

We would like to emphasize that our systematic derivation relegates spacetime covariance to a lesser role than that of electric-magnetic symmetry.   This feature, previously encountered in several other instances, might convey an important lesson for the investigations of more general ``hidden symmetries" that extend electric-magnetic duality,  such as $E_{10}$ or $E_{11}$ \cite{Hidden}.  

Although our discussion has covered an ample realm of cases of physical interest, they were all concerned with $p$-forms, which are totally antisymmetric tensors.  There are important cases, which were not covered herein and which will be addressed in a forthcoming publication \cite{BHtoappear2}.  They are linearized gravity \cite{Henneaux:2004jw} and higher spin fields \cite{Bunster:2006rt}.  In those cases, the electric and magnetic ``superpotentials" have mixed symmetries.

\section*{Acknowledgments} 
We thank John Schwarz for kind clarifying comments. M. H.  gratefully acknowledges support from the Alexander von Humboldt Foundation through a Humboldt Research Award and support from the ERC through the ``SyDuGraM" Advanced Grant.  The Centro de Estudios Cient\'{\i}ficos (CECS) is funded by the Chilean Government through the Centers of Excellence Base Financing Program of Conicyt. The work of M. H. is also partially supported by IISN - Belgium (conventions 4.4511.06 and 4.4514.08), by the Belgian Federal Science Policy Office through the Interuniversity Attraction Pole P6/11 and by the ``Communaut\'e Fran\c{c}aise de Belgique" through the ARC program.  We extend our thanks to the referees for their very pertinent observations.

\end{document}